\documentclass{nature}
\usepackage{times}
\usepackage{graphicx}
\usepackage{amscd}
\usepackage{amsmath,amssymb,amsthm,mathrsfs,amsfonts,dsfont}
\usepackage{subfigure, epsfig}
\usepackage{braket}
\usepackage{bm}
\usepackage{enumerate}
\usepackage{graphicx}
\usepackage{epsfig}
\usepackage{epstopdf}
\usepackage{subfigure}
\usepackage{color}

\title{Quantum key distribution over 658 km fiber with distributed vibration sensing}

\begin{document}
\maketitle
\author
{
 Jiu-Peng Chen,$^{1,2,3,\dagger}$ Chi Zhang,$^{1,2,3,\dagger}$ Yang Liu,$^{3}$ Cong Jiang,$^{3}$ Dong-Feng Zhao,$^{1}$ Wei-Jun Zhang,$^{4}$ Fa-Xi Chen,$^{3}$ Hao Li,$^{4}$ Li-Xing You,$^{4}$ Zhen Wang,$^{4}$ Yang Chen,$^{1}$ Xiang-Bin Wang,$^{2,3,5}$ Qiang Zhang,$^{1,2,3}$ Jian-Wei Pan$^{1,2}$
}

\begin{affiliations}
\item Hefei National Laboratory for Physical Sciences at Microscale and Department of Modern Physics, University of Science and Technology of China, Hefei 230026, China
\item CAS Center for Excellence and Synergetic Innovation Center in Quantum Information and Quantum Physics, University of Science and Technology of China, Hefei 230026, China
\item Jinan Institute of Quantum Technology, Jinan, Shandong 250101, China
\item State Key Laboratory of Functional Materials for Informatics, Shanghai Institute of Microsystem and Information Technology, Chinese Academy of Sciences, Shanghai 200050, China
\item State Key Laboratory of Low Dimensional Quantum Physics, Department of Physics, Tsinghua University, Beijing 100084, China\\
$\dagger$ These authors contributed equally: Jiu-Peng Chen, Chi Zhang.

\end{affiliations}

\begin{abstract}
Twin-field quantum key distribution (TF-QKD)~\cite{lucamarini2018overcoming} promises ultra-long secure key distribution which surpasses the rate distance limit~\cite{PLOB2017} and can reduce the number of the trusted nodes in long-haul quantum network~\cite{natureyuao}. Tremendous efforts have been made towards implementation of TF-QKD, among which, the secure key with finite size analysis can distribute  more than 500 km in the lab~\cite{fang2019surpassing,chen2020sending,pittaluga2021600} and in the field~\cite{chen2021twin}. Here, we demonstrate the sending-or-not-sending TF-QKD~\cite{wang2018sns} experimentally, achieving a secure key distribution with finite size analysis over 658 km ultra-low-loss optical fiber, improve the secure distance record by around 100 km. Meanwhile, in a TF-QKD system, any phase fluctuation due to temperature variation and ambient variation during the channel must be recorded and compensated, and all these phase information can then be utilized to sense the channel vibration perturbations. With our QKD system, we recovered the external vibrational perturbations on the fiber generated by an artificial vibroseis and successfully located the perturbation position with a resolution better than 1 km. Our results not only set a new distance record of QKD, but also demonstrate that the redundant information of TF-QKD can be used for remote sensing of the channel vibration, which can find applications in earthquake detection and landslide monitoring besides secure communication.
\end{abstract}

\section*{Introduction}
Quantum key distribution (QKD)~\cite{bennett1984quantum,gisin2002quantum,gisin2007quantum,scarani2009security,xu2020secure,pirandola2020advances} offers the theoretical provable way to distribute secure keys. However, the channel loss is an inevitable barrier for long distance QKD since a quantum signal cannot be amplified. For a transmission $\eta$, the theoretical upper bound of the secure key rate is limited to 1.44 $\eta$, known as the Pirandola-Laurenza-Ottaviani-Bianchi (PLOB) bound~\cite{PLOB2017}. This upper bound is valid for all the repeaterless QKD protocols which include the commonly decoy-state based BB84~\cite{hwang2003quantum,wang2005beating,lo2005decoy}, and the measurement-device-independent QKD (MDIQKD)~\cite{lo2012measurement,braunstein2012side} which closes all security loopholes of measurement devices. Without the practical quantum repeater, an intermediate solution to achieve long haul QKD network is to set several trusted relay nodes. Although the trusted relay networks are successfully demonstrated in the field~\cite{natureyuao}, the increased number of trusted relays might increase the security risk and raise the cost.

Different from the traditional QKD protocols, the twin-field QKD (TF-QKD)~\cite{lucamarini2018overcoming} improves the secure key rate scaling to $\sqrt\eta$ without using quantum memory. This may provides a solution to reach a longer distance and to reduce the number of trusted relays. Recently, the feasibility of distributing secure keys over long distance is proved experimentally~\cite{minder2019experimental,wang2019beating,liu2019exp,zhong2019proof,chen2020sending,fang2019surpassing,liu2021field,pittaluga2021600,chen2021twin}. Notably, with full security analysis considering the finite size effect, experimental demonstration of sending or not sending TF-QKD (SNS-TF-QKD)~\cite{wang2018sns} is realized with record long-distance of more than 500 km in lab~\cite{chen2020sending,pittaluga2021600} and in field ~\cite{chen2021twin}. In order to achieve a secure final key, one needs to overcome the challenging problem finite key effects with a relatively small data size. In the case of not considering data finite size, one can even obtain a positive key rate at a distance of 600 km~\cite{pittaluga2021600}.

Realizing TF-QKD is challenging, because the protocols requires phase sensitive single-photon interference. Any phase differences, caused by laser wavelength differences or channel fiber vibration may reduce the interference visibility. Techniques such as time-frequency metrology~\cite{liu2019exp,chen2020sending,chen2021twin}, optical phase locking loop (OPLL)~\cite{wang2019beating,minder2019experimental,pittaluga2021600} have been developed to eliminate the wavelength difference; real-time~\cite{wang2019beating,pittaluga2021600} or post-processing~\cite{liu2019exp,fang2019surpassing,chen2020sending,liu2021field,chen2021twin} compensation have been developed to eliminate the fast fiber vibration.

Besides supporting TF-QKD, the obtained information in fast phase compensation actually reflects the real-time phase variation of the transmitted light in the optical fiber. This information can also be utilized to detect the vibrational perturbation in the channel. As such, the redundant information obtained in an installed TF-QKD system might be used as a fiber-optic sensor to detect critical vibrations in the channel. Different from the well-known distributed acoustic fiber sensing (DAS)~\cite{mestayer2011field,lindsey2017fiber,Nathaniel2019} technique, the phase-tracking method used in TF-QKD analyzes the transmitted light, not the back-scattered light. This technique is similar to the phase-based frequency metrology interferometric technique~\cite{numatathermal2004,foreman2007remote,drosteoptical2013,lisdatclock2016,delvatest2017,GiuseppeMarra2018}, making it possible to achieve an ultra-long vibrational sensing length. With this interferometric method, the measured phase signal will be the result of integration of perturbations along the whole fiber. Fortunately, by using the simutaneous bidirectional phase-tracking~\cite{GiuseppeMarra2018}, it is possible to identify the perturbation location by cross-correlating the time difference between the signals of Alice and Bob.

Here, we demonstrated SNS-TF-QKD~\cite{wang2018sns} experimentally through a 658 km ultra-low loss optical fiber with a total loss of 106 dB. The secure key rate is $9.22\times 10^{-10}$ per pulse after collecting 27.8 hours data for considering finite key size effect in security analysis. Meanwhile, we insert an artificially vibroseis in the channel to generate specific vibration signals. With the same TF-QKD experimental setup, we recovered the vibration signals generated by the vibroseis. Further, by cross-correlating the vibrational signals at the two users, we successfully located the vibroseis to a 1 km precision over the 500 km frequency locking fibers, which is, as far as we know the longest reported distance~\cite{GiuseppeMarra2018,yan2021unidirectional}.

\section*{Protocol}
We adopt the 4-intensity sending-or-not-sending (SNS) protocol~\cite{wang2018sns} in the experiment, and apply the actively odd parity pairing (AOPP)~\cite{jiang2019unconditional,xu2019general,jiang2021composable} method for post data processing to achieve a higher key rate. In this protocol, Alice and Bob send pulse pairs to the detection station, Charlie. For the received pulse pair, if Charlie announces there is only one detector clicks, it is a one-detector heralded events. In each side of Alice and Bob, there are three weak coherent state (WCS) sources with intensities $\mu_1$, $\mu_2$, 0 in the decoy windows, and two sources with intensity $\mu_z$ and 0 in the signal windows. The one-detector heralded events in the decoy windows are used to perform decoy state analysis, and the corresponding bits of the one-detector heralded events in the signal windows are used to extract the final keys. See Supplemental Information for details of the theory calculation method.

\section*{Experiment}
\begin{figure*}[htb]
\centering
\resizebox{14cm}{!}{\includegraphics{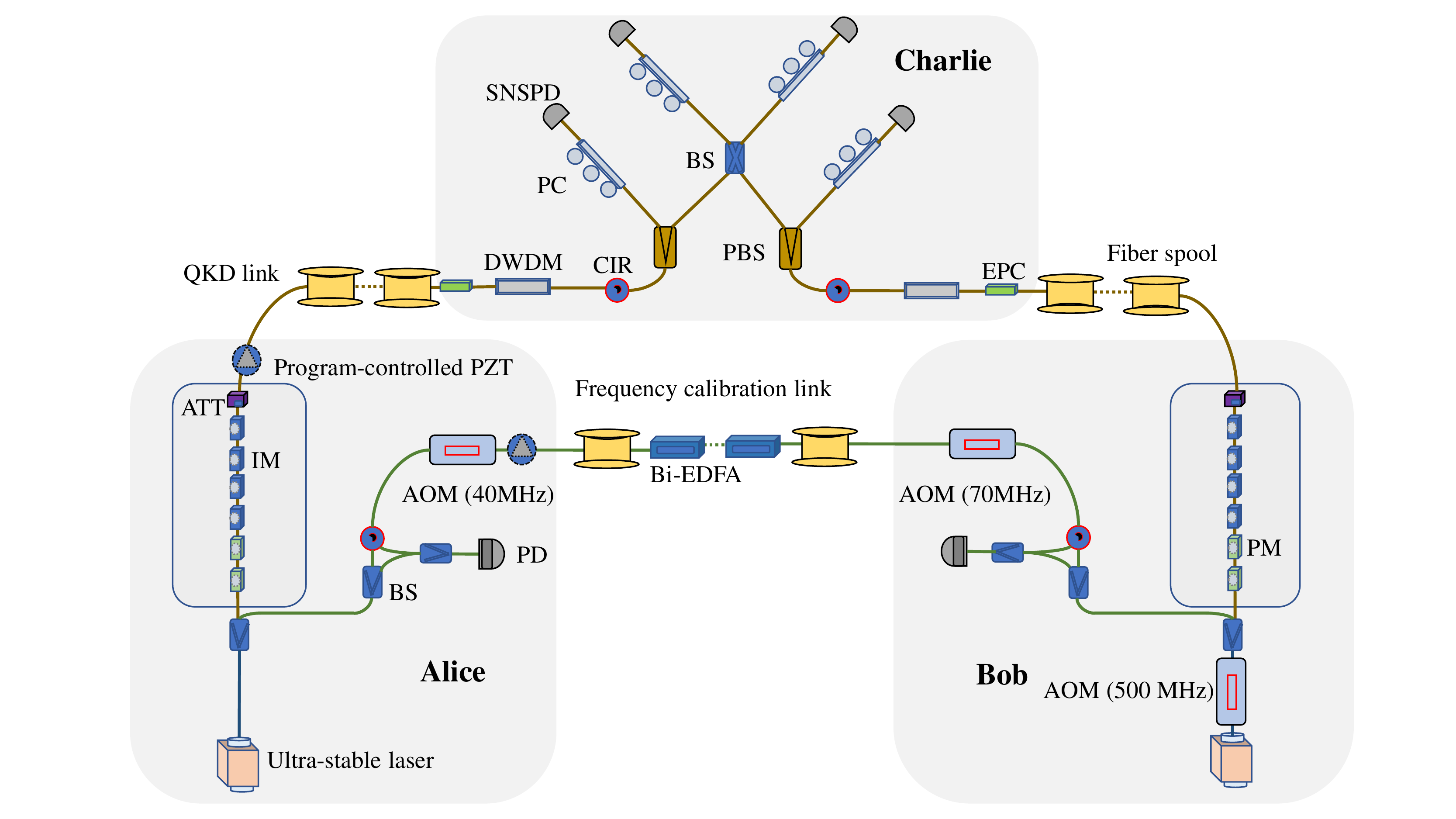}}
\caption{\textbf{Schematic of experimental setup.} \baselineskip12pt In Alice's (Bob's) lab, a seed laser is locked to a ultra-low-expansion (ULE) glass cavity to achieve sub-Hz linewidth by using the Pound-Drever-Hall (PDH)~\cite{pound1946electronic,drever1983laser} technique. The ultra-stable laser is split to two parts, one is used to calibrate the wavelength difference between the local and remote users, and the other is used as the QKD source. In the wavelength calibration part, the light passes an acoustic-optic modulator (AOM) with 40 MHz (70 MHz) fixed carrier frequency in Alice (Bob), to filter the noise in the channel. Then the light is sent to Bob (Alice) for heterodyne detection with a photodiode (PD). An AOM with 500 MHz $\pm$ 5 MHz adjustable carrier frequency is inserted at Bob to eliminate the frequency difference of Alice's and Bob's light source. The wavelength calibration light travels through about 500 km SMF-28 spools between Alice and Bob. Bidirectional erbium-doped fiber amplifiers (bi-EDFAs) are inserted every 50 km to maintain the power of the transmitted light. In the QKD part, the light is modulated with phase modulators (PMs) and intensity modulators (IMs) and attenuate to single photon level with an attenuator (ATT), to generate the quantum signals with the phase reference signals. The light is finally sent to Charlie via 329.3 km and 329.4 km ultra-low loss fiber spools (658.7 km) for detection. Charlie uses a Dense Wavelength Division Multiplexer (DWDM), a circulator (CIR) to filter the noises before the polarization beam splitter (PBS) and the beam splitter (BS). The interference results are detected by superconducting nanowire single-photon detectors (SNSPDs). Additionally, the fiber stretchers are insert in the QKD channel and the wavelength calibration channel, as the artificial vibroseis. EPC: electric polarization controller, PC: polarization controller.}
\label{Fig:seisim-setup}
\end{figure*}

The experimental setup is shown in Fig~\ref{Fig:seisim-setup}. Alice and Bob use two independent ultra-stable lasers of which the relative frequency difference is eliminated. The light is modulated to a pattern that the single-photon-level quantum signal pulses are time multiplexed with strong phase reference pulses. In each period of 1 $\mu$s, 100 signal pulses with 240 ps pulse duration are sent within the first 400 ns, and the phase reference pulses are sent in the following 600 ns. The signals from Alice and Bob are sent to Charlie through 329.3 km and 329.4 km (658.7 km in total) ultra-low loss fiber spools with a transmission of 52.9 dB and 53.1 dB (106 dB in total). After interference at Charlie's beam splitter (BS), the signals are detected by two superconducting nanowire single photon detectors (SNSPDs), and recorded by a time tagger. (See Supplemental Information for details of the experimental setup.)

The key to realize SNS-TF-QKD is the stable single photon interference, which requires the Alice's and Bob's lasers are locked to the same frequency. We solve this problem by adapting the frequency metrology technology. First, the linewidth of Alice's (Bob's) seed laser is suppressed to sub-Hz by locking it to an ultra-low-expansion (ULE) glass cavities using the Pound-Drever-Hall (PDH) technique~\cite{pound1946electronic,drever1983laser}. After PDH locking, the relative frequency drift rate is smaller than 0.1 Hz/s. Then, a 500 km fiber spools is set between Alice and Bob to calibrate the relative frequency difference of their lasers. Heterodyne detecting is performed to monitor the frequency difference on both sides, and an acoustic-optic modulator (AOM) with adjustable carrier frequency is inserted at Bob's output to compensate the frequency difference. In this frequency calibration setup, additional AOMs with fixed carrier frequency are inserted at both ends of the link to eliminate the channel reflection in frequency domain, and to shift transmitted laser wavelengths for heterodyne detection.

To achieve single photon interference, the relative phase between Alice and Bob should also be compensated. We adapted strong phase reference pulses time multiplexed with the signal pulses to detect and compensate the fast phase fluctuation. The phase drift is attributed to the change of the refractive index and length of fiber. By neglecting the effect of the long-term temperature drift, the phase difference at the interference is mainly contributed by the accumulation of mechanical perturbations such as vibration and sound through the long fiber channel. Now that the fast phase fluctuation can be compensated in TF-QKD, the phase perturbations induced by vibration through the channel can be derived straightforwardly. In other words, we can regard the TF-QKD system as a sensing equipment to detect vibrations in the channel.

Besides for the use of obtaining the relative frequency difference information of the two laser sources, the frequency calibration link can also be used to detect vibration. The vibration on the fiber will induce the frequency/phase perturbations of the transmitted light by affecting the refractive index or the length of the fiber. This frequency/phase perturbations of the transmitted light is measured by detecting the phase of the radio frequency (RF) signal of heterodyne. The relative delay of the detected frequency/phase perturbations signals in Alice and Bob indicates the distance differences between Alice and Bob to the vibroseis. Thus, the position of the vibroseis can be located by cross-correlating the detected phase perturbations signals of Alice and Bob. To investigate the vibration perturbations, we inserted program-controlled PZT vibration generators in the quantum channel and frequency calibration channel, as shown in Fig~\ref{Fig:seisim-setup}.  (See Supplemental Materials for details of the vibration test methods.)

\section*{Results}
In the experiment, we first explore the longest possible TF-QKD distribution distance with our setup. A 658 km G.652 ultra-low loss fiber with a total loss of 106 dB is used as the quantum channel, which is 0.161 dB/km on average, including the connections. The component loss and insertion loss is optimized to 1.3 dB in Charlie. Then, we adopted high performance SNSPDs with a detection efficiency of 82\% and an effective dark count rate of 4 Hz to detect the interference, and set a time gate of 0.3 ns to suppress noise. The final noise is optimized to be $6\times10^{-9}$ per pulse, about 80\% of which is from re-Rayleigh scattering~\cite{chen2020sending}.

The main results of our experiment are shown in Tab.~\ref{Tab:Parameters}. In about 27.8 hours, a total of $1.007\times 10^{13}$ signals are sent at the 100 MHz effective system frequency, yielding $5.28\times10^6$ valid detections. We observe a quantum phase flip error rate (QBER) in X basis of around 5\%, with a base-line error rate of around 2.8\%. The bit-flip error rate in Z basis is 26.29\% before actively odd parity pairing (AOPP)~\cite{xu2019general,jiang2019unconditional}, mainly contributed by the optimized ``sending" probability of 0.2717 in Z basis. The QBER in Z basis decreases to 2.12\% after AOPP, while the phase error rate increases to 13.36\% and the survived bits drops to 244731 from 558729.

The secure key rate is then calculated following Eq.~\eqref{eq:KeyRate}, considering the finite data size effect~\cite{jiang2019unconditional,jiang2020zigzag}:
\begin{equation}\label{eq:KeyRate}
\begin{split}
 R=&\frac{1}{N_t}\{ n_1^\prime[1-H(e_1^{ph})]-fn_t^\prime H(E_Z)\\
 &-2\log_2\frac{2}{\varepsilon_{cor}}-2\log_2\frac{1}{\sqrt{2}\varepsilon_{PA}\hat{\varepsilon}}\},\\
\end{split}
\end{equation}
where $R$ is the final key rate, $n_1^\prime$, $e_1^{ph}$,  $n_t^\prime$ and $E_Z$ are the number of untagged-bits, the phase-flip error rate, the number of survived bits, and the bit-flip error rate of untagged-bits after AOPP. $f = 1.16$ is the error correction efficiency. $N_{t}$ is the total number of signal pulses, $\varepsilon_{cor}=1\times 10^{-10}$ and $\varepsilon_{PA}=1\times 10^{-10}$ are the failure probability of error correction process and privacy amplification process, $\hat{\varepsilon}=1\times 10^{-10}$ is the coefficient of the chain rules of smooth min- and max- entropy.

\begin{figure*}[htb]
\centering
\resizebox{12cm}{!}{\includegraphics{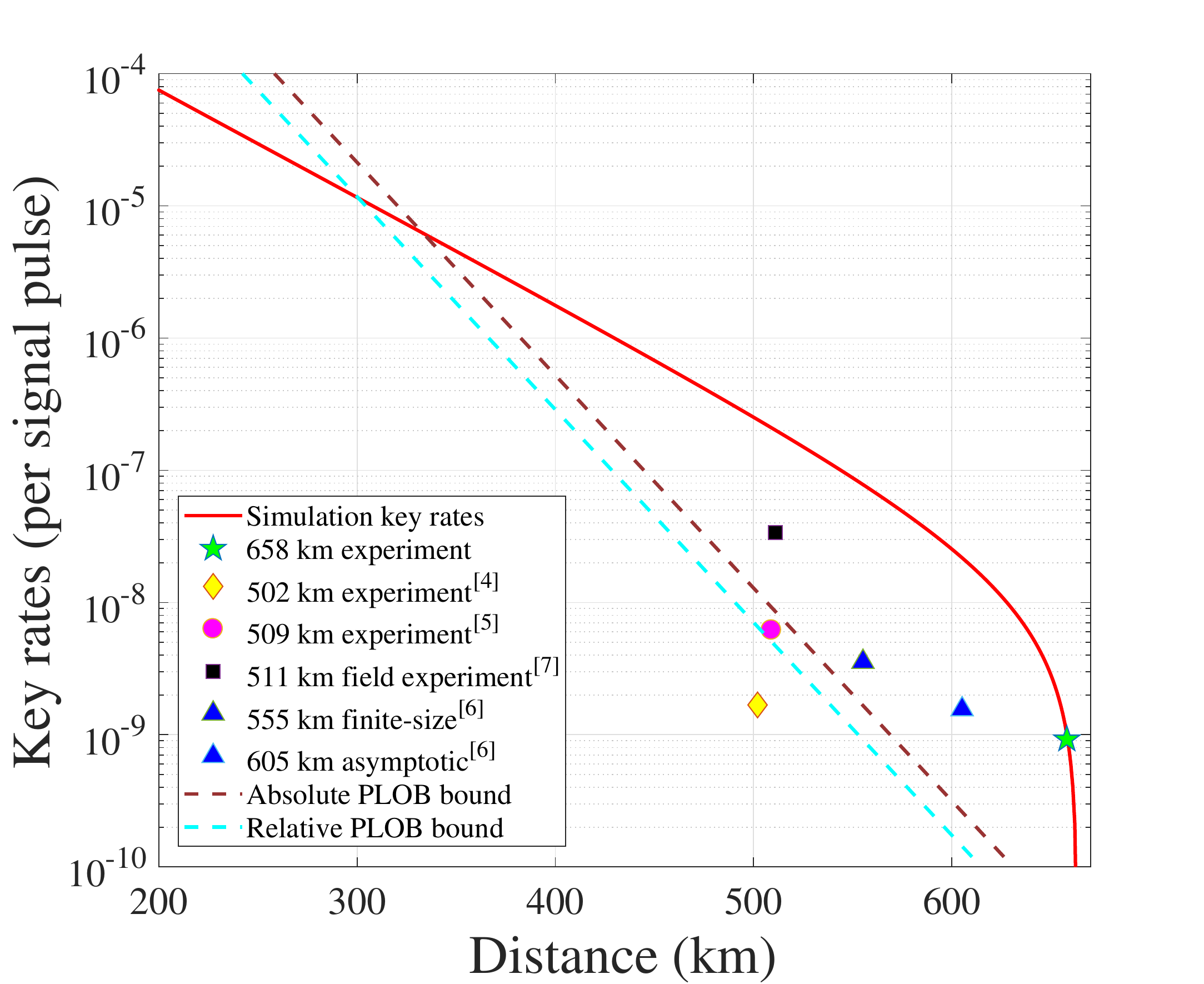}}
\caption{\textbf{Secure key rates of the SNS-TF-QKD experiment.} The green star indicates the experimental result over  658 km ULL fibers, with the secure key rate of $R=9.22\times10^{-10}$. The yellow diamond, purple circle and blue triangle indicate the experimental results of ref.~\cite{fang2019surpassing}, ref.~\cite{chen2020sending} and ref.~\cite{pittaluga2021600} in the lab. The black square indicates the experimental result of ref.~\cite{chen2021twin} in field. The red curve is the simulation result with the experimental parameters. The brown dotted line and cyan dotted line show the absolute and relative PLOB bound~\cite{PLOB2017}.}
\label{Fig:key-rate}
\end{figure*}

The final secure key rate is $R=9.22\times10^{-10}$, which is about 0.092 bit per second (bps) considering 100 MHz effective system frequency. We summarize our theoretical simulation and experimental result in Fig.~\ref{Fig:key-rate}. The obtained secure key rate here is more than one order of magnitude higher than the absolute PLOB bound. (See Supplemental Information for details of experimental parameters and results.)
\begin{table*}[htb]
\centering
\caption{Experimental results. The parameters $n_1^\prime$ is the number of untagged-bits, $e_1^{ph}$ is the phase-flip error rate, $n_t^\prime$ is the number of survived bits, $E_Z$ is the bit-flip error rate of untagged-bits after AOPP, and $N_{t}$ is the total number of signal pulses.}
\begin{tabular}{c|cc}
\hline \hline
$n_t^\prime$  & 558729 \\
$n_1^\prime$  & 244731  \\
$e_1^{ph}$   & 13.36\%  \\
$E_Z$   & 2.12\%\\
\hline
$N_{t}$	 & $1.007\times 10^{13}$ \\
Valid detections &$5.28\times10^{6}$\\
Key rate			& $9.22\times10^{-10}$ \\ \hline \hline
\end{tabular}
\label{Tab:Parameters}
\end{table*}

\begin{figure*}[htb]
\centering
\resizebox{12cm}{!}{\includegraphics{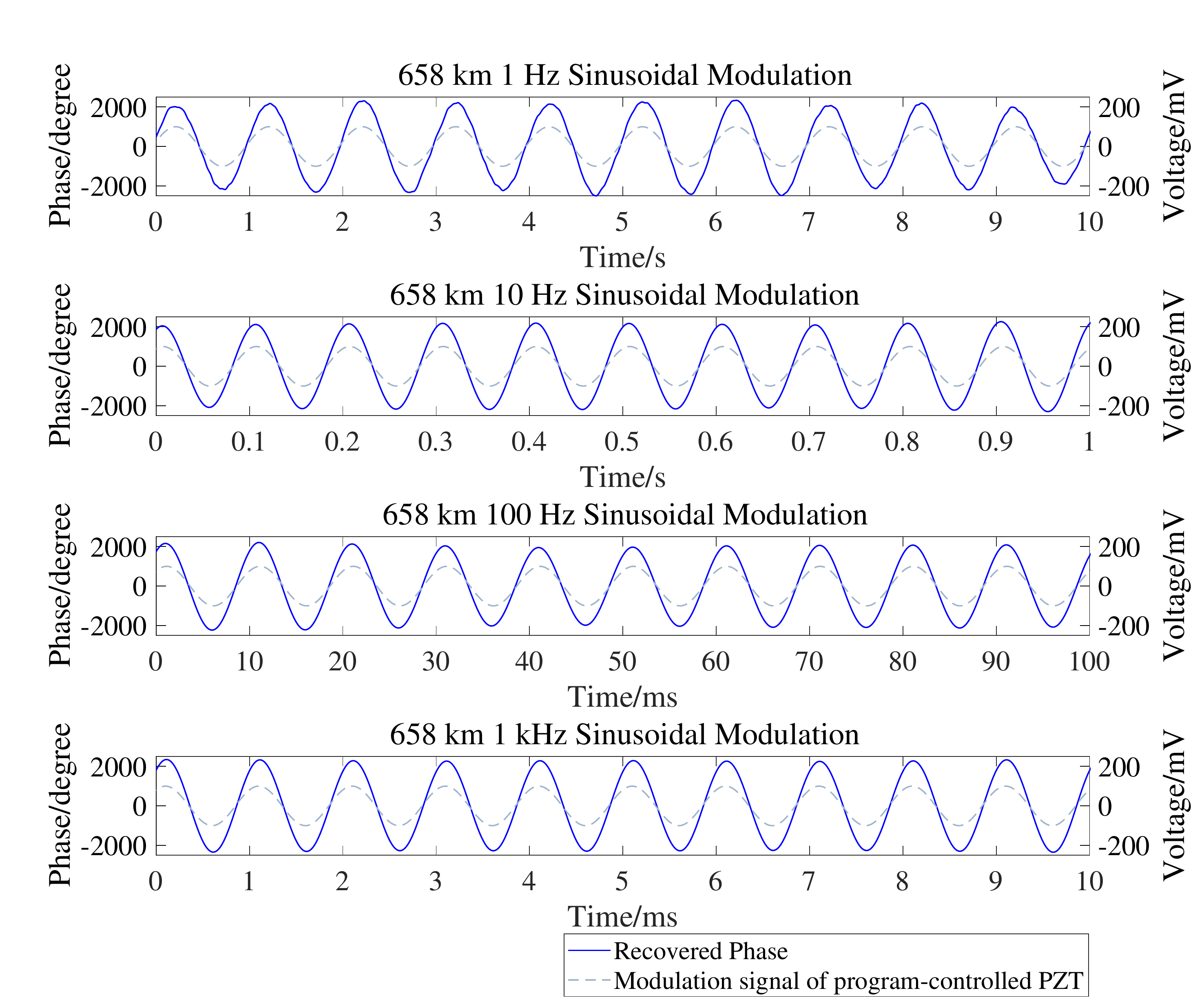}}
\caption{\textbf{Vibration test results via QKD link.} The program-controlled PZT is inserted in the QKD channel which is set to a length of 658 km. The modulation signal of program-controlled PZT is set to sinusoidal wave with selected frequencies of 1 Hz, 10 Hz, 100 Hz, and 1 kHz, respectively. The blue curve indicates the recovered relative phase variation signal. The gray dotted line indicates the modulation signal of program-controlled PZT.}
\label{Fig:seis-QKD}
\end{figure*}

Next, we modulate the PZT vibration generators with fixed frequencies to simulate the vibration perturbations in the channel. In the case the PZT vibration takes place in the 658 km quantum channel, the phase drift is recovered by consequently calculating the relative phase difference with the phase reference pulses. We set the modulation to sinusoidal signal with selected frequencies of 1 Hz, 10 Hz, 100 Hz, and 1 kHz, respectively, which is the frequency range of interest in seismic and acoustic wave sensing. In the time domain, the recovered phase variation perfectly matches the active modulation signal, i.e., the externally applied vibration on the fiber. The driving signals and the recovered phases are plotted in Fig.~\ref{Fig:seis-QKD}.

\begin{figure*}[htb]
\centering
\resizebox{12cm}{!}{\includegraphics{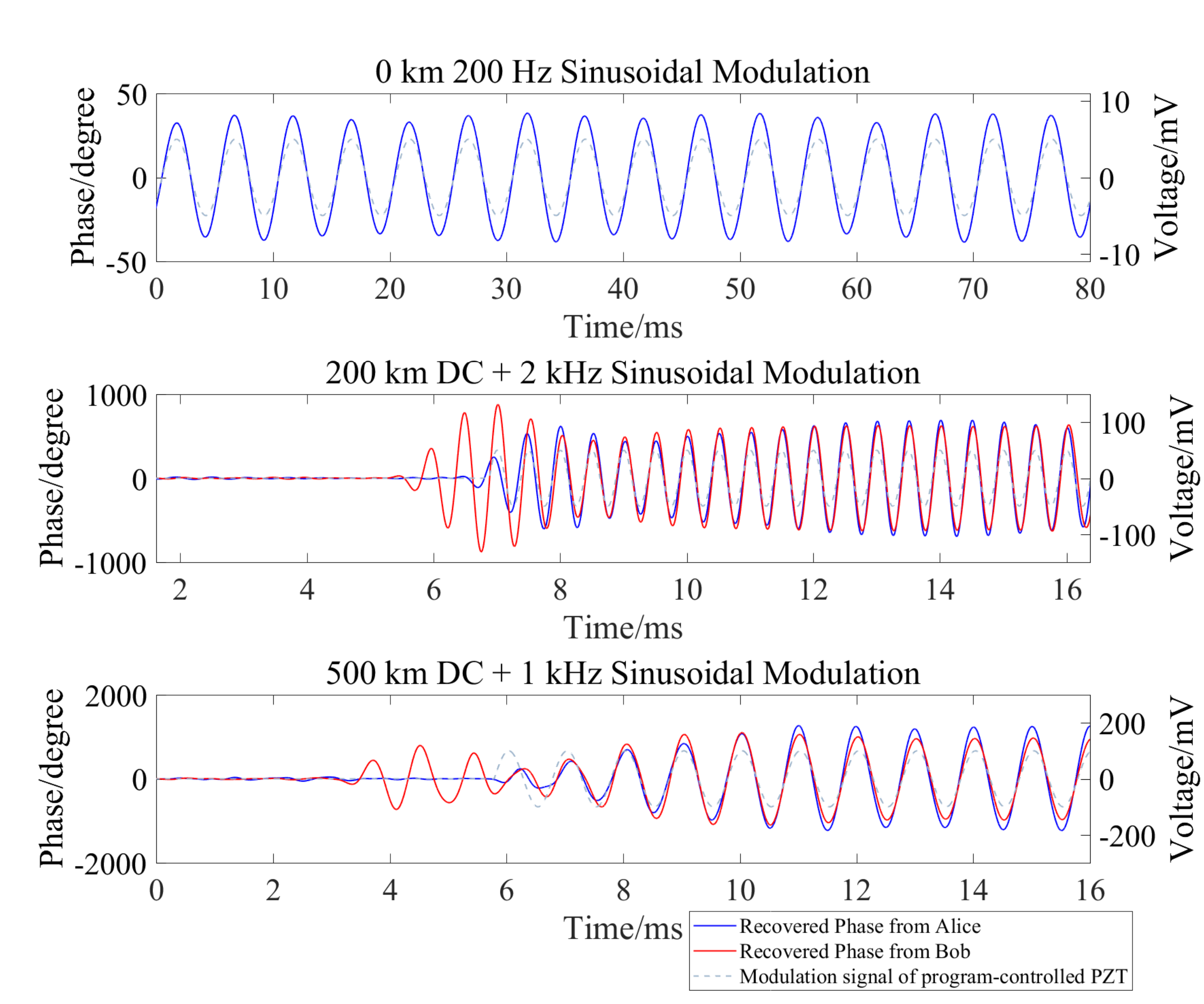}}
\caption{\textbf{Vibration test results via frequency calibration link.} The program-controlled PZT is inserted in the frequency calibration channel which is set to a length of 0 km, 200 km and 500 km. The corresponding modulation signal of PZT is set to 200 Hz sinusoidal wave, direct current (DC) combined with 2 kHz sinusoidal wave, and DC combined with 1 kHz sinusoidal wave, respectively. The blue curve and red curve indicate the recovered phase variation signals of Alice and Bob. The gray dotted line indicates the modulation signal of program-controlled PZT.}
\label{Fig:seis-cal}
\end{figure*}

In the case the PZT vibration takes place in the frequency calibration channel, we set the channel length to 0 km, 200 km, and 500 km, respectively, and install the vibroseis at Alice with different vibration frequencies. Here, the vibration signal is recovered by electronically decoding phase perturbations of the RF signal of heterodyne. As shown in Fig.~\ref{Fig:seis-cal}, the recovered phase variation shows a frequency and waveform exactly the same as that of the driving signal in all the fiber lengths. The vibroseis position is measured by calculating the relative time delay of the vibration signals at Alice and Bob. For the case of a fiber length of 500 km and the vibroseis installed at Alice as an example, the relative time delay between the Alice and Bob's signals is measued as 2.531 ms by corss-correlation. By adopting the speed of light in the fiber to be $2.0\times10^{8} m/s$, this yields a location of the vibtaion cource at 502.6 km away from Bob. Similarly, the vibroseis is located to 200.0 km away from Bob, with the relative delay to be 1.000 ms, when the fiber length is set to 200 km and the vibroseis installed at Alice. The precision of location is better than 1 km, which is mainly limited by the sampling rate of our phase measurement.

\section*{Conclusion}
In conclusion, we demonstrated SNS-TF-QKD over 658 km ultra-low loss optical fiber spools experimentally, achieving a secure key rate of $9.22\times 10^{-10}$ per pulse with the finite key size effect considered. We recover 1 Hz - 1kHz vibration perturbations on the fiber with the phase reference and the frequency locking channel, and locate the vibroseis with a precision better than 1 km over the 500 km fiber spools. Our work provides a proof of principle that the TF-QKD architecture is able to be used for ultra-long distance vibration sensing, while distributing secure keys. We expect that the developed techniques may expand the application of QKD networks, specifically in the field of earthquake detection, landslide monitoring and highway traffic monitoring, etc, where a distributed seismic detection is necessary.\bigskip

\section*{References}
\bibliography{SeismicSNSTFQKD}

\begin{addendum}
\item[Acknowledgements]This work was supported by the National Key R\&D Program of China (Grants No. 2017YFA0303900, 2017YFA0304000, 2020YFA0309800), the National Natural Science Foundation of China (T2125010 ), the Chinese Academy of Sciences (CAS), Key R\&D Plan of Shandong Province (Grant No. 2019JZZY010205, 2020CXGC010105), and Anhui Initiative in Quantum Information Technologies.
\item[Competing interests] The authors declare no competing interests.
\end{addendum}

\clearpage

\clearpage

\clearpage

\clearpage

\end{document}